\documentclass{epl}
\usepackage{amssymb,graphicx,psfrag}
\title{Optical conductivity of d-wave superconductors}
\shorttitle{Optical conductivity of d-wave superconductors}
\author{Bal\'azs D\'ora\inst{1} \and Kazumi Maki\inst{2} 
\and Attila Virosztek\inst{1,3}}
\institute{
\inst{1} Department of Physics, Technical University of Budapest, H-1521 Budapest, Hungary \\
\inst{2} Department of Physics and Astronomy, University of Southern California, Los Angeles
CA 90089-0484, USA \\
\inst{3} Research Institute for Solid State Physics and Optics, P.O.Box 49,
H-1525 Budapest, Hungary}
\shortauthor{Bal\'azs D\'ora \etal}


\date{}

\begin{document}

\maketitle

\begin{abstract}
We study theoretically the optical conductivity of d-wave superconductors like
in high temperature cuprates in the presence of impurities. We limit ourselves
at $T=0K$ and focus on the frequency dependence of both $\sigma_1(\omega)$ and 
$\omega\sigma_2(\omega)$ for $\omega\lesssim 2\Delta$. When the impurity scattering
 is in the unitary limit, we find a peak in $\sigma_1(\omega)$
with $\omega/\Delta\simeq 0.1\sim 0.5$, which may account for the peak seen by Basov et al.
in $Zn$-substituted $YBCO$.
\end{abstract}

\section{Introduction}

Now d$_{x^2-y^2}$-wave superconductivity is well established in both hole-doped and 
electron-doped high temperature cuprate superconductors \cite{dsc1,dsc2,dsc3,dsc4,dsc5}.
 Further superconductivity
in organic conductors in $\kappa-(ET)_2$ salts appears to be of d-wave 
\cite{et_2-1,et_2-2}. In spite
of these developments, the optical conductivity in d-wave superconductors appears
to be not well understood. For example Hirschfeld et al. \cite{hirsch1} considered the microwave
conductivity in d-wave superconductors in the presence of impurity in the unitary limit.
However they were more interested in the temperature dependence rather than the
frequency dependence. Similarly Sun and Maki \cite{maki1} and Graf et al. 
\cite{graf} studied only some
aspect of the optical conductivity. On the other hand, a recent infrared conductivity
in $Zn$-substituted $YBCO$ exhibits a peak around
$\omega/\Delta\simeq0.1\sim 0.2$, which cannot 
be accounted for the above theories \cite{basov}.
The object of this paper is to study the in plane optical conductivity focusing on
the frequency dependence at $T=0K$.
As a model we take d-wave superconductors (with gap function $\Delta({\bf k})=\Delta\cos(2\phi)$,
$\phi$ is the in plane angle measured from the $k_x$ direction) in the presence of impurities in the unitary
and Born limit. We assume that the former describes the $Zn$ impurity while the latter 
the $Ni$ impurity \cite{maki2}.
In the following we summarize what we needed for the model, then we calculate 
$\sigma_1(\omega)$ and $\omega\sigma_2(\omega)$ for impurities in the unitary
and the Born limit for several impurity concentrations. Indeed, for the impurity
in the unitary limit $\sigma_1(\omega)$ develops a peak around $\omega/\Delta\simeq0.1\sim 0.2$,
somewhat similar to the one observed in $Zn$-substituted $YBCO$. On the other hand, 
$\sigma_1(\omega)$ in the Born limit develops a broad bump around $\omega/\Delta\sim1.5$. 

\begin{figure}[h]
\psfrag{x}[t][b][1][0]{$\Gamma/\Gamma_c$}
\psfrag{y}[b][t][1][0]{$\Delta(0,\Gamma)/\Delta_{00}$, $T_c/T_{c0}$ and 
$N(0,\Gamma)/N_0$}
\twofigures[width=7cm,height=7cm]{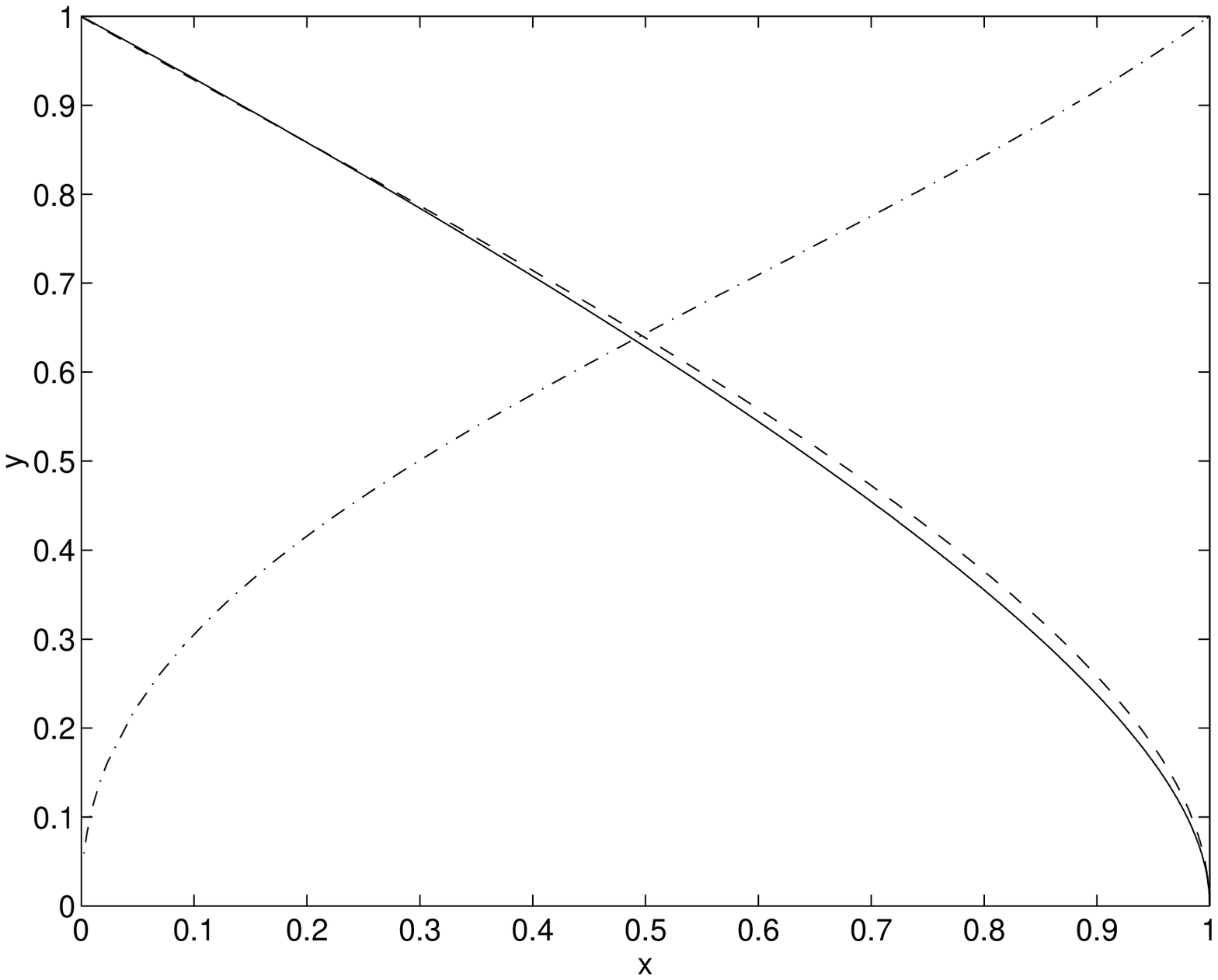}{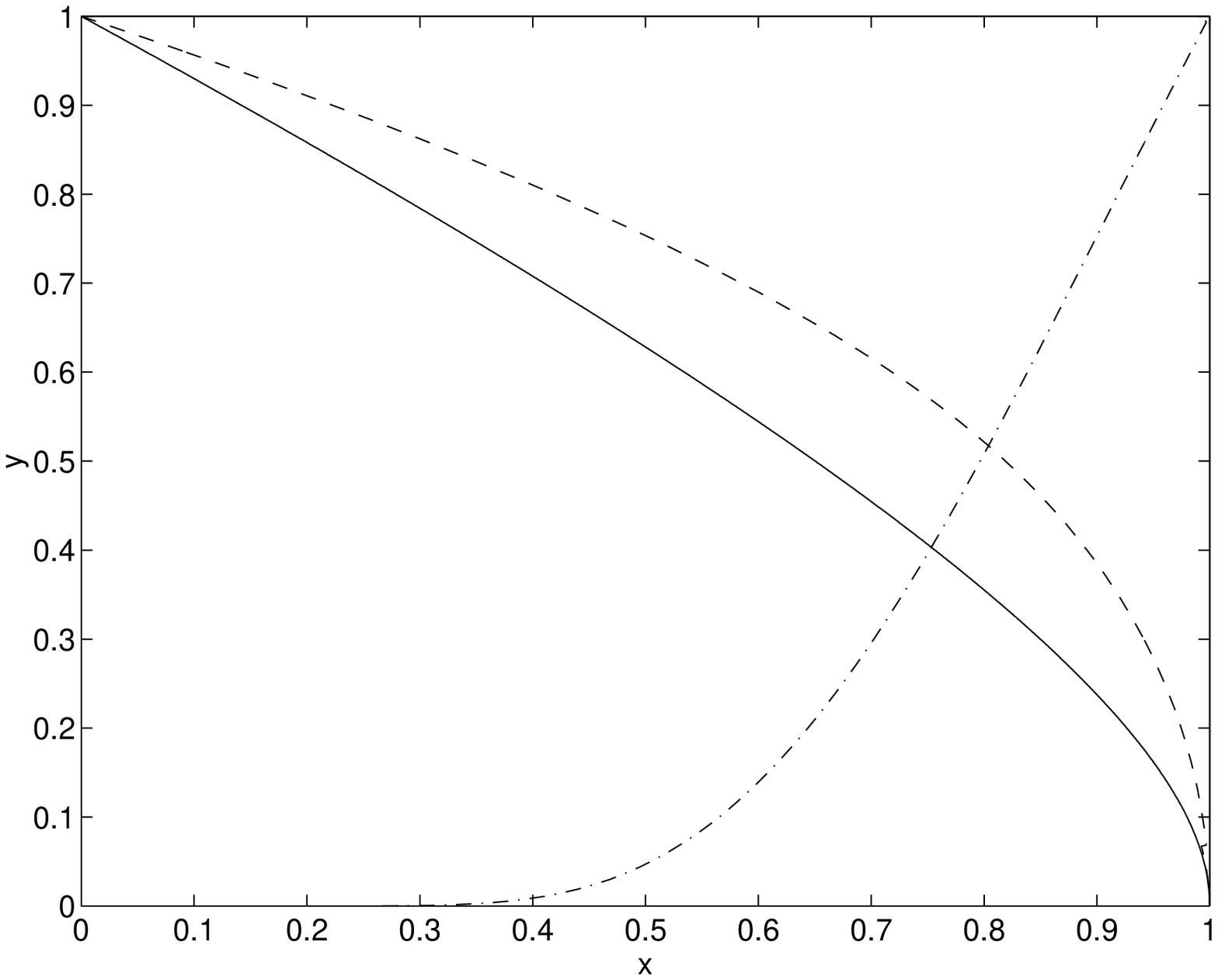}

\caption{$\Delta(0,\Gamma)/\Delta_{00}$ (dashed line), $T_c/T_{c0}$ (solid
line) and $N(0,\Gamma)/N_0$ (dashed-dotted line) are shown as a function of
$\Gamma/\Gamma_c$ in the unitary limit.}
\label{fig:dtnu}
\caption{$\Delta(0,\Gamma)/\Delta_{00}$ (dashed line), $T_c/T_{c0}$ (solid
line) and $N(0,\Gamma)/N_0$ (dashed-dotted line) are shown as a function of
$\Gamma/\Gamma_c$ in the Born limit.}
\label{fig:dtnb}
\end{figure}

The effect of impurity scattering is incorporated by renormalizing the frequency
 in the quasi-particle Green's function \cite{maki3,maki4}:
\begin{eqnarray}
\frac{\omega}{\Delta}=u+\alpha\frac\pi 2
\frac{\sqrt{1-u^2}}{uK\left(\frac{1}{\sqrt{1-u^2}}\right)}, \\
\frac{\omega}{\Delta}=u-\alpha\frac2\pi
\frac{u}{\sqrt{1-u^2}}K\left(\frac{1}{\sqrt{1-u^2}}\right)
\end{eqnarray}
for the unitary and the Born limit, respectively, where
$u=\tilde{\omega}/\Delta$, $\alpha=\Gamma/\Delta$, $K(z)$ is the complete
elliptic integral of the first kind and $\Gamma$ is the scattering rate.
Making use of the gap equation the superconducting transition temperature
($T_c$) in the presence of impurity is given:
\begin{equation}
-\ln\left(\frac{T_c}{T_{c0}}\right)=\Psi\left(\frac12+\frac{\Gamma}{2\pi
T_c}\right)-\Psi\left(\frac12\right),
\end{equation}
where $T_{c0}$ is the one in the absence of impurity and $\Psi(z)$ is the
di-gamma function. This is the same as the Abrikosov-Gor'kov formula for
s-wave superconductors in the presence of magnetic impurities \cite{mester}. Also it is
independent of whether the impurity is in the unitary limit or in the Born
limit. Further we note the AG formula is universal in the sense that it applies to all unconventional superconductors (p-wave, d-wave, f-wave etc.) and independent on whether the scattering is in the unitary or the Born limit. Now at $T=0K$ the gap equation yields \cite{maki2,maki3,maki4}
\begin{eqnarray}
-\ln\left(\frac{\Delta(0,\Gamma)}{\Delta_{00}}\right)=2 \langle f^2
\textmd{arcsinh}\left(\frac{C_0}{f}\right)\rangle- \nonumber \\
-2\frac\Gamma\Delta\int_{C_0}^\infty dx\frac{1}{x^2}\left(1-\frac{E}{K}\right)\left((1+x^2)\frac
E K -x^2\right),
\end{eqnarray}
and
\begin{eqnarray}
-\ln\left(\frac{\Delta(0,\Gamma)}{\Delta_{00}}\right)=2 \langle f^2
\textmd{arcsinh}\left(\frac{C_0}{f}\right)\rangle+ \nonumber \\
+2\left(\frac 2\pi\right)^2\frac\Gamma\Delta\int_{C_0}^\infty dx(K-E)\left(E-\frac{x^2}{1+x^2}K\right)
\end{eqnarray}
for the unitary limit and the Born limit, respectively and $C_0$ is given
by 
\begin{equation}
C_0^2=\frac{\pi\Gamma}{2\Delta}\sqrt{1+C_0^2}\left[
K\left(\frac{1}{\sqrt{1+C_0^2}}\right)\right]^{-1}
\end{equation}
and 
\begin{equation}
\sqrt{1+C_0^2}=\frac{2\Gamma}{\pi\Delta}
K\left(\frac{1}{\sqrt{1+C_0^2}}\right)
\end{equation}
for the unitary limit and the Born limit, respectively, and
$f=\cos(2\phi)$, $\Delta_{00}=\Delta(0,0)$, $\langle \dots \rangle$ means average of $\phi$ and
$K=K(1/\sqrt{1+x^2})$ and $E=E(1/\sqrt{1+x^2})$. Finally the residual density
of states (i.e. the density of states on the Fermi surface) is given
by
\begin{eqnarray}
\frac{N(0,\Gamma)}{N_0}=\frac2\pi
\frac{C_0}{\sqrt{1+C_0^2}}K\left(\frac{1}{\sqrt{1+C_0^2}}\right)=\left\{
\begin{array}{lc}
\Gamma/(\Delta C_0) & \textmd{unitary limit}\\
\Delta C_0/\Gamma   & \textmd{Born limit}.
\end{array}
\right.
\end{eqnarray}
We show in fig. \ref{fig:dtnu} and  fig. \ref{fig:dtnb}  $T_c/T_{c0}$, $\Delta(0,\Gamma)/\Delta_{00}$
and $N(0,\Gamma)/N_0$ versus $\Gamma/\Gamma_c$ for the unitary and Born
limit respectively, and $\Gamma_c=0.8819T_{c0}$.

\section{Quasi-particle density of states and optical conductivity}

\begin{figure}
\psfrag{x}[t][b][1][0]{$E/\Delta(\Gamma)$}
\psfrag{y}[b][t][1][0]{$N(E,\Gamma)/N_0$}
\twofigures[width=7cm,height=7cm]{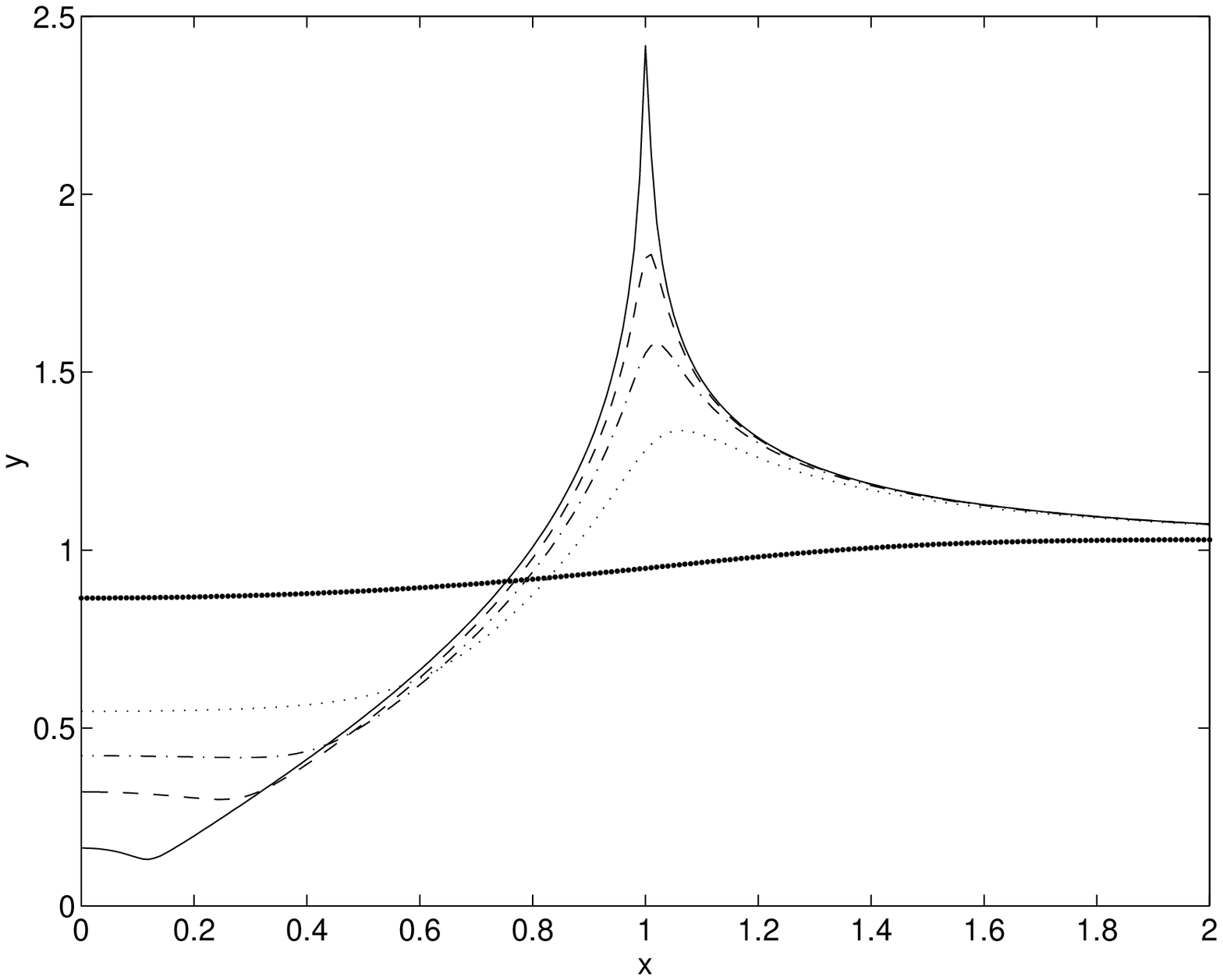}{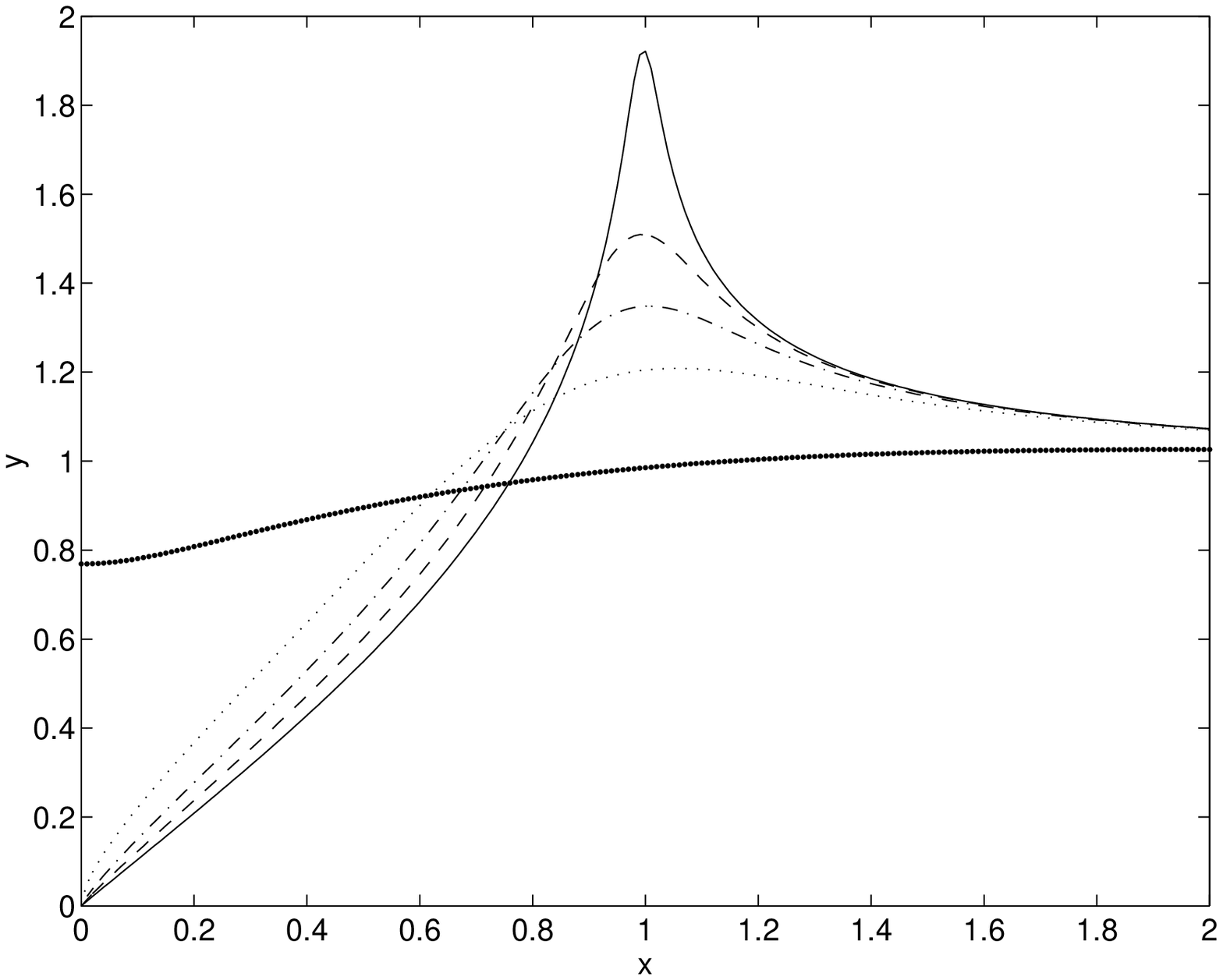}

\caption{The density of states is shown for different
$\Gamma/\Delta(\Gamma)$ in the unitary limit: 0.01 (thin solid line), 0.02
(dashed line), 0.1 (dashed-dotted line), 0.2 (thin dotted line) and 1 (thick solid line).}
\label{fig:dosu}
\caption{The density of states is shown for different
$\Gamma/\Delta(\Gamma)$ in the Born limit: 0.01 (thin solid line), 0.02
(dashed line), 0.1 (dashed-dotted line), 0.2 (thin dotted line) and 1 (thick solid line).}
\label{fig:dosb}
\end{figure}

The quasi-particle density of states in the presence of impurities is given
by
\begin{equation}
\frac{N(0,\Gamma)}{N_0}=\textmd{Re}\left\langle\frac{u}{\sqrt{u^2-f^2}}\right\rangle.
\end{equation}
We show the quasi-particle density of states for $\Gamma/\Delta=0.01$,
$0.05$, $0.1$, $0.2$ and $1$ for the unitary and Born limit in
fig. \ref{fig:dosu}  and \ref{fig:dosb}, respectively. 
These figures are consistent with the earlier results in \cite{graf,maki2,maki3,maki4}.
In particular in the Born limit there is little density of states at $E=0$
until $\Gamma/\Gamma_c\ge 0.5$. This result is consistent with tunneling
conduction data from $Ni$ substituted $Bi2212$ \cite{Nibizmut}. 

\begin{figure}
\psfrag{x}[t][b][1][0]{$\Gamma/\Gamma_c$}
\psfrag{y}[b][t][1][0]{$\sigma_1(0)m\pi\Delta(0,0)/e^2n$}
\onefigure[width=7cm,height=7cm]{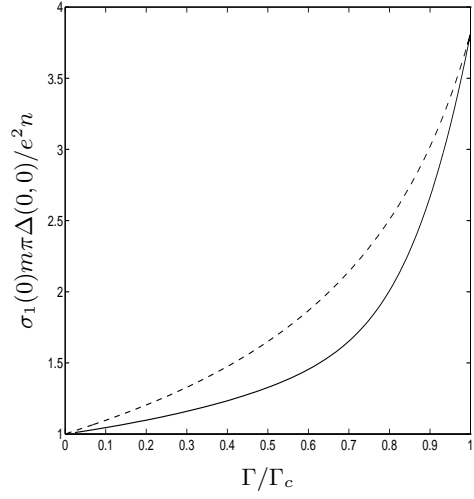}
\caption{The DC conductivity is shown as a function of $\Gamma/\Gamma_c$
in the unitary (dashed line) and the Born (solid line) limit.}
\label{fig:dcvez}
\end{figure}

Now the optical conductivity is expressed as
$\sigma(\omega)=\sigma_1(\omega)+i\sigma_2(\omega)$:
 
\begin{eqnarray}
\sigma_1(\omega)=-\frac{e^2n}{\omega m \pi \Delta} \textmd{Re}I(\omega),\\
\omega\sigma_2(\omega)=-\frac{e^2n}{m \pi \Delta} \textmd{Im}\left(I(\omega)+2
\int_{-\infty}^\infty\frac{1}{e^{\beta x}+1}F(u(x),u(x-\omega))dx\right),
\end{eqnarray}
where 
\begin{eqnarray}
I(\omega)=\int_{-\infty}^\infty \frac 1 2 \left(\tanh\left(\frac{\beta x}{2}\right)-\tanh\left(
\frac{\beta(x+\omega)}{2}\right)\right)\times \nonumber \\
\times(F(u(x+\omega),\overline{u}(x))-F(u(x+\omega),u(x)))
dx
\end{eqnarray}
and
\begin{equation}
F(u,u^\prime)=\frac{1}{u^\prime-u}\left(\frac{u^\prime}{\sqrt{1-{u^\prime}^2}}K\left(\frac{1}{\sqrt
{1-{u^\prime}^2}}\right)-\frac{u}{\sqrt{1-u^2}}K\left(\frac{1}{\sqrt{1-u^2}}\right)\right).
\end{equation}
It is worth noting, that the imaginary part of the integral of the second $F$ function in
$I(\omega)$ is zero. In evaluating $F(u,u^\prime)$, we included self
energy corrections from impurities, but the vertex corrections vanished in
the long wavelength limit because of s-wave scattering \cite{p-wave,SDW}.
At zero frequency, $\sigma_1(\omega)$ reduces to:
\begin{equation}
\sigma_1(0)=\frac{e^2n}{m\pi
\Delta(0,\Gamma)}\frac{E\left(\frac{1}{\sqrt{1+C_0^2}}\right)}{\sqrt{1+C_0^2}},
\end{equation}
which is valid in both limits. $E(z)$ is the complete elliptic integral of
 the second kind. At $\Gamma_c$, $\sigma_1(0)$ reaches $\pi/0.8244$ times
 its pure value. See fig. \ref{fig:dcvez}! Note the vertical axis in fig. \ref{fig:dcvez} can be rewritten
as $\sigma_1(0)/\sigma_n=lim_{T\rightarrow 0}\kappa/\kappa_n$, where $\kappa$ is the thermal conductivity
and $\sigma_n$ and $\kappa_n$ are the conductivity and the thermal conductivity in the normal state. In the vortex
state of d-wave superconductors the Wiedeman-Franz law still holds \cite{maki1,maki2,graf2}. Then fig. \ref{fig:dcvez}
shows the deviation from Lee's universality relation \cite{Lee}. Indeed this deviation from the universality is 
verified later in $Zn$-substituted $YBCO$ by measuring the low temperature thermal conductivity \cite{ybco}.

\begin{figure}
\psfrag{x}[t][b][1][0]{$\omega/\Delta(0,0)$}
\psfrag{y}[b][t][1][0]{$\sigma_1(\omega)m\pi\Delta(0,0)/e^2n$}
\twofigures[width=7cm,height=7cm]{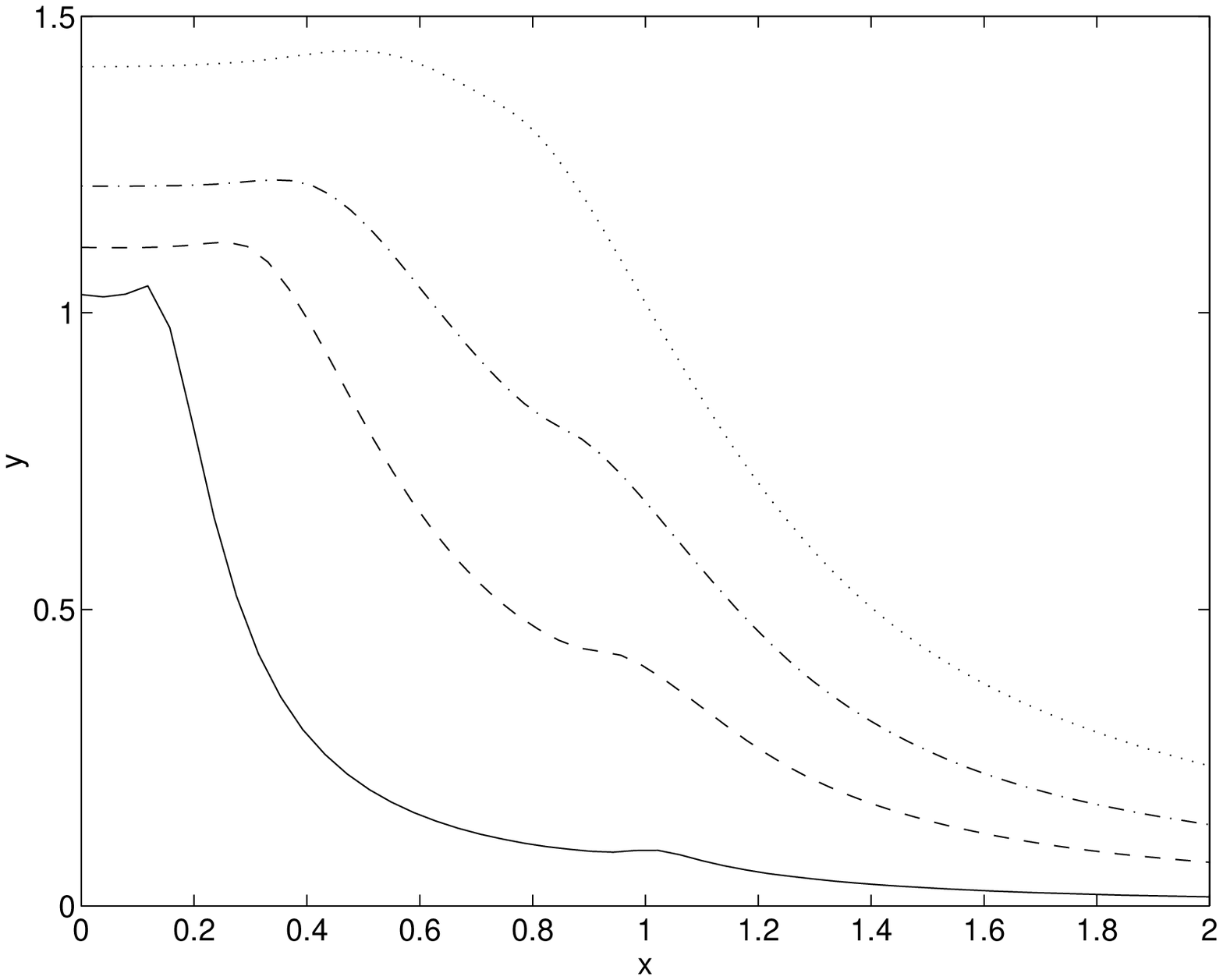}{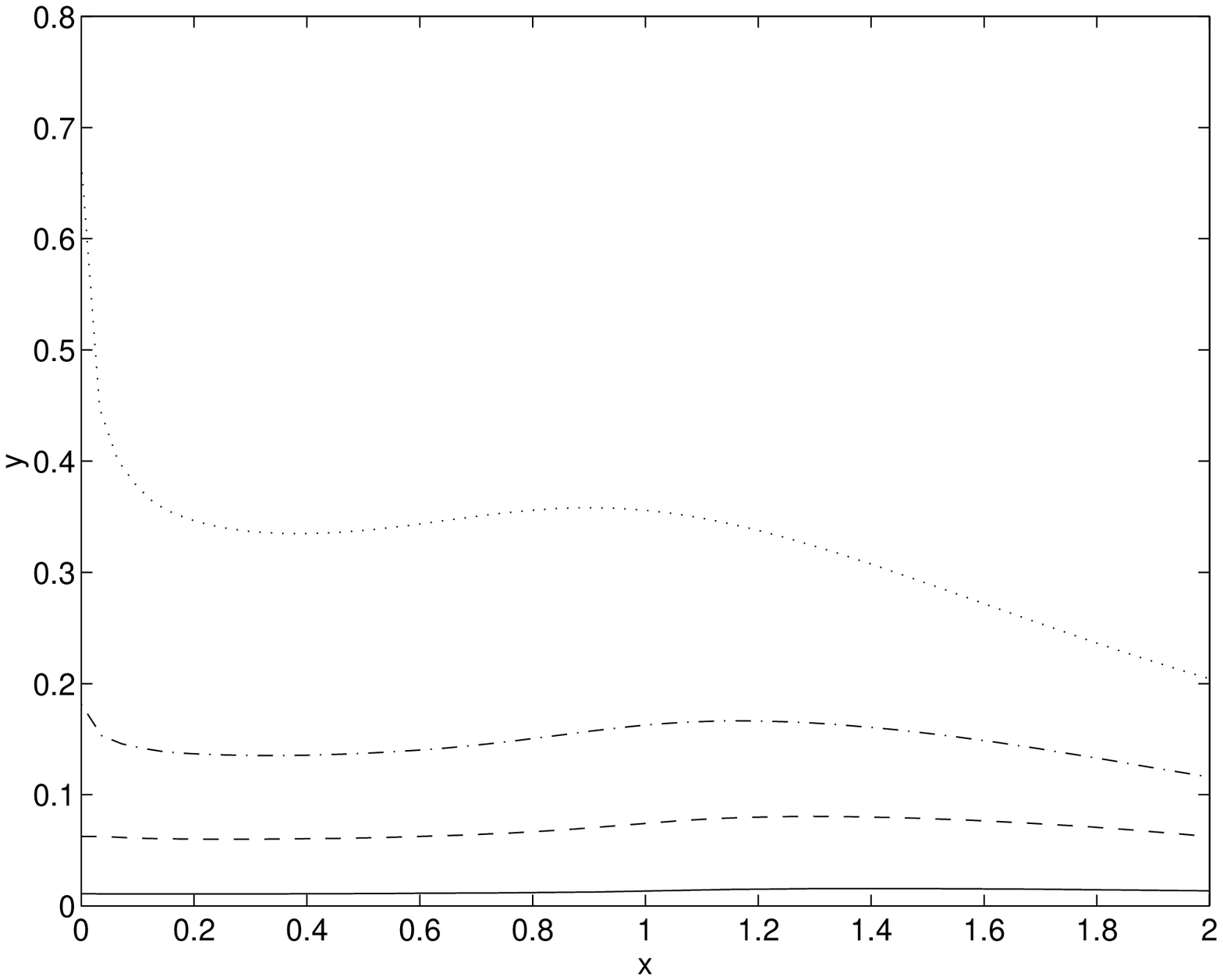}

\caption{Real part of the optical conductivity is plotted for different
$\Gamma/\Delta(\Gamma)$ in the unitary limit: 0.01 (thin solid line), 0.02
(dashed line), 0.1 (dashed-dotted line) and 0.2 (thin dotted line).}
\label{fig:revezu}
\caption{Real part of the optical conductivity is plotted for different
$\Gamma/\Delta(\Gamma)$ in the Born limit: 0.01 (thin solid line), 0.02
(dashed line), 0.1 (dashed-dotted line) and 0.2 (thin dotted line).}
\label{fig:revezb}
\end{figure}

\begin{figure}
\psfrag{x}[t][b][1][0]{$\omega/\Delta(0,0)$}
\psfrag{y}[b][t][1][0]{$\omega\sigma_2(\omega)m\pi/e^2n$}
\twofigures[width=7cm,height=7cm]{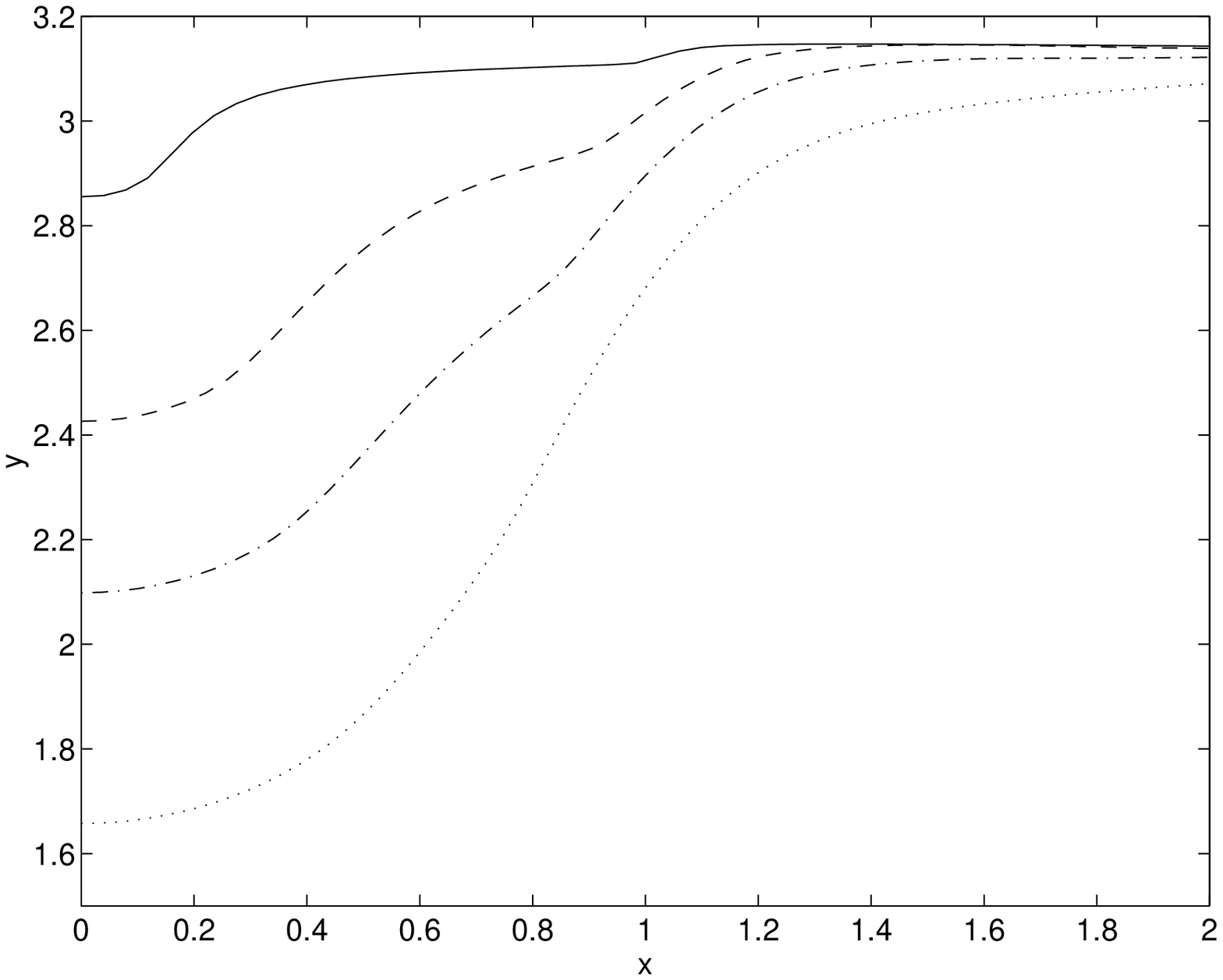}{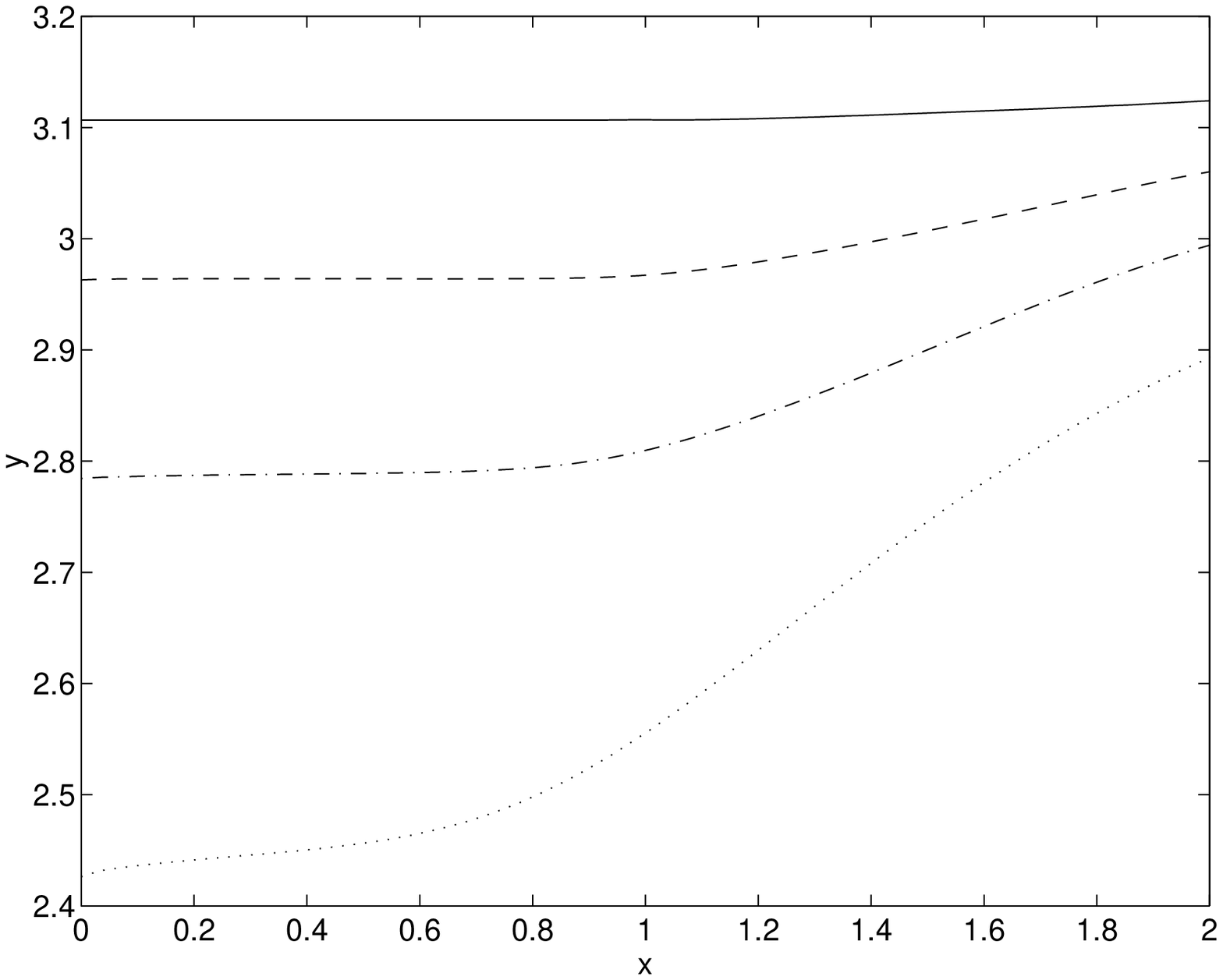}

\caption{$\omega\sigma_2(\omega)$ is shown for different
$\Gamma/\Delta(\Gamma)$ in the unitary limit: 0.01 (thin solid line), 0.02
(dashed line), 0.1 (dashed-dotted line) and 0.2 (thin dotted line).}
\label{fig:imvezu}
\caption{$\omega\sigma_2(\omega)$ is shown for different
$\Gamma/\Delta(\Gamma)$ in the Born limit: 0.01 (thin solid line), 0.02
(dashed line), 0.1 (dashed-dotted line) and 0.2 (thin dotted line).}
\label{fig:imvezb}
\end{figure}

 In fig. \ref{fig:revezu} and \ref{fig:revezb}, we show $\sigma_1(\omega)$ at $T=0K$ for the previously
investigated scattering rates in the unitary and Born limit. In the unitary
limit $\sigma_1(\omega)$ develops a small peak around
$\omega/\Delta\simeq0.1\sim0.2$ for $\Gamma/\Delta<0.1$, this may be
related to the small peak seen in $Zn$ substituted $Y(124)$ \cite{basov}. On the other
hand in the Born limit $\sigma_1(\omega)$ initially drops monotonously
with increasing $\omega$, then develops a broad bump around
$\omega/\Delta\sim 1.5$. In either case there is no clear feature around
$\omega/\Delta\sim 2$. This implies perhaps the single particle rather than
the pair breaking scattering dominates $\sigma_1(\omega)$.

In fig. \ref{fig:imvezu} and \ref{fig:imvezb}, we show $\omega\sigma_2(\omega)$ at $T=0K$ for the unitary
and the Born limit. We note that
\begin{equation}
\lim_{\omega\rightarrow
0}\omega\sigma_2(\omega)=\frac{e^2n}{m}\rho_s(0,\Gamma),
\end{equation}
where $\rho_s(0,\Gamma)$ is the superfluid density. Further
$\lim_{\omega\rightarrow \infty}\omega\sigma_2(\omega)=e^2n/m$ independent
of $\Gamma$ and of whether in the unitary limit or the Born limit. This
means that $\omega\sigma_2(\omega)$ should have a dip at $\omega=0$ as seen
in these figures as well as in \cite{maki1}.
In the Born limit, there is a small peak at $\omega/\Delta=1$ for small
scatterers, which is unobservable in our figure due to its scale.
Also in the unitary limit this dip like structure appears to develop steps.
In a future study we shall discuss the temperature dependence of these
quantities.

\section{Conclusion}
We find a simple closed form expression of the complex conductivity for
d-wave superconductors which applies for both the unitary and the Born
limit. In the unitary limit $\sigma_1(\omega)$ exhibits a small peak around
$\omega/\Delta\simeq 0.1\sim0.2$ which may describe a similar feature
observed experimentally. Finally we discover that a dip in
$\omega\sigma_2(\omega)$ at $\omega=0$ is the universal feature of the
unconventional superconductors, which has been overlooked until now.

Finally the present model will apply as well to the superconductivity in
$Sr_2RuO_4$, if the superconductivity is one of 2D-f-wave states
($\Delta({\bf k})\sim e^{\pm i\phi}\cos(2\phi)$, $e^{\pm i\phi}\sin(2\phi)$ without the
vertex renormalization and $e^{\pm i\phi}\cos(k_zc)$) considered by Hasegawa et
al. \cite{2Df1,2Df2,2Df3,2Df4}. However, both anisotropy in the upper
critical field in a planar magnetic field \cite{critfield1,critfield2} and the magnetic thermal
conductivity in a planar magnetic field \cite{hovezetes1,hovezetes2} indicate extremely small 
angular dependence ($\sim 3\%$). Clearly these data show that $\Delta({\bf k})\sim e^{\pm i\phi}\cos(2\phi)$ 
and $e^{\pm i\phi}\sin(2\phi)$ are incompatible with this observation. Therefore at this moment we are left with 
$\Delta({\bf k})\sim e^{\pm i\phi}\cos(k_zc)$ as the only candidate. However this suggests a rather 
strong interlayer spin coupling, which is very puzzling.

We thank Dimitri Basov for useful discussion on the optical data of $Zn$-
substituted $YBCO$ and providing us with some of his unpublished data. We are
benefited also from useful discussions with and help of Hyekyung Won. One of
the authors (B. D.) gratefully acknowledges the hospitality at the University
of Southern California, Los Angeles, where part of this work was done. This work
was supported by the Hungarian National Research Fund under grant numbers
OTKA T032162 and T029877, and by the Ministry of Education under grant number
FKFP 0029/1999.

\bibliographystyle{unsrt}
\bibliography{theory}
\nocite{*}

\end{document}